# COCCIA LAB

*To discover the causes of social, economic and technological change*

# How do scientific disciplines evolve in applied sciences? The properties of scientific fission and ambidextrous scientific drivers

Mario COCCIA

CNR -- National Research Council of Italy
&
Yale University

# How do scientific disciplines evolve in applied sciences? The properties of scientific fission and ambidextrous scientific drivers

*Mario Coccia*[1]

CNR -- National Research Council of Italy & Yale University
CNR -- National Research Council of Italy
Collegio Carlo Alberto, Via Real Collegio, 30-10024 Moncalieri (Torino, Italy)
Yale School of Medicine
310 Cedar Street, Lauder Hall, New Haven, CT 06510, USA
*E*-mail: mario.coccia@cnr.it

Mario Coccia ORCID: http://orcid.org/0000-0003-1957-6731

## Abstract

The evolution of science is made possible when experimental results are compared with expectations from theory and are consistent. In this context, experimental physics, as applied science, plays a vital role for the progress of science in society. The experimental physics is a discipline where physics scholars have an intensive laboratory experience that concentrates on experiments for substantiating and/or challenging established and/or new theories in physics. No studies to date allows us to explain the endogenous processes that support the evolution of scientific disciplines and emergence of new scientific fields in applied sciences of physics. In fact, one of the fundamental questions in science is how scientific disciplines evolve and sustain progress in society. This study confronts this question here by investigating the evolution of experimental physics to explain and generalize, whenever possible, some characteristics of the dynamics of applied sciences. Empirical analysis suggests a number of properties about the evolution of experimental physics and in general of applied sciences, such as: *a) scientific fission,* the evolution of scientific disciplines generates a process of division into two or more research fields that evolve as autonomous entities over time; *b) ambidextrous drivers of science*, the evolution of science *via* scientific fission is due to scientific discoveries or new technologies; *c) new driving research fields*, the drivers of scientific disciplines are new research fields rather than old ones (e.g., three scientific fields with a high scientific production in experimental physics are emerged after 1950s); *d) science driven by development of general purpose technologies,* the evolution of experimental physics and applied sciences is due to the convergence of experimental and theoretical branches of physics associated with the development of computer, information systems and applied computational science (e.g., computer simulation). Results also reveal that average duration of the upwave of scientific production in scientific fields supporting experimental physics is about 80 years. Overall, then, this study begins the process of clarifying and generalizing, as far as possible, some characteristics of the evolutionary dynamics of scientific disciplines that can lay a foundation for the development of comprehensive properties explaining the evolution of science as a whole for supporting fruitful research policy implications directed to advancement of science and technological progress in society.

---

[1]   Acknowledgement. I gratefully acknowledge financial support from National Research Council of Italy–Direzione Generale Relazioni Internazionali for funding this research project developed at Yale University in 2019 (grant-cnr n. 62489-2018). The author declares that he has no relevant or material financial interests that relate to the research discussed in this paper.

**Introduction and goal of this investigation**

The goal of this paper is to analyze the evolution of a vital scientific discipline in physics, the experimental physics, and to suggest empirical characteristics and properties of endogenous processes of the evolution of experimental physics that can explain and generalize the evolutionary dynamics of research fields in applied sciences over time and space.

This study is part of a large body of research on the evolution of science to explain how scientific disciplines emerge, evolve and decline in human society (Coccia, 2018, Coccia and Wang, 2016; Sun et al., 2013). The evolution of science and scientific fields has been explored with different scientific perspectives[2]. Many studies have investigated the structure of science, using maps that show scientific landscape to identify major fields of science, their size, similarity, and interconnectedness (Börner and Scharnhorst, 2009; Boyack et al., 2005; Clark, 1987; Simonton, 2004). Other studies endeavor to explain the role of social interactions in shaping the dynamics of science and the emergence of new disciplines (Börner et al., 2011; Tijssen, 2010; Sun et al., 2013: Van Raan, 2000)[3], the evolution and convergence between research fields considering international research collaboration (Coccia and Bozeman, 2016; Coccia and Wang, 2016), etc.

However, the characteristics of the evolution of research fields remain still ambiguous and ill-defined for explaining the general evolution of science for appropriate research policy. Stimulated by these fundamental problems in the field of social study of science and knowledge, this paper endeavors to clarify the following question concerning the evolution of scientific disciplines:

- Which are the endogenous processes of the evolution of experimental physics and in general of scientific disciplines in applied sciences?

The literature about this question is rather scarce but these topics are critical to science and society for understanding the evolution of scientific fields and designing a research policy directed to support science advances and new technology for wellbeing in society (Coccia, 2005, 2014, 2019; De Solla Price, 1986; Kitcher, 2001; Latour, 1987; Storer, 1967; Stephan and Levin, 1992; Sun et al. 2013). In short, there is need for much more detailed research to explain the evolution of research fields and find general properties. This study confronts the question just mentioned by developing an inductive analysis, which describes evolutionary characteristics and properties of experimental physics, a vital scientific discipline in applied sciences. Results of this study may afford an interesting opening into the exploration of empirical properties that explain and generalize, whenever possible, the evolution of scientific disciplines in applied sciences and, in general, of endogenous processes of scientific development. In order to position this study in a manner that displays similarities and differences with existing approaches, next section begins by reviewing accepted theoretical frameworks of scientific development in social studies of science.

---

[2] cf., Adams, 2012; Ávila-Robinson et al., 2019; Coccia and Bozeman, 2016; Freedman, 1960; Kuhn, 1962; Lakatos, 1968, 1978; Lee and Bozeman, 2005; Merton, 1957, 1968; Souzanchi Kashani and Roshani, 2019; Stephan, 1996; Zhou et al., 2019.
[3] cf., Boyack, 2004; Boyack et al., 2005; Fanelli and Glänzel, 2013; Simonton, 2002; Small, 1999; Smith et al., 2000; Sun et al., 2013.

**Patterns of scientific development**

Seidman (1987) states that: "science is an organized and collective activity (p. 131) …scientific development occurs in a dynamic relation to the encompassing social context (p. 134) …. Society is constitutive of science not merely in the sense of forming a normative context enhancing or impeding scientific rationality, but in that it informs the very processes of inquiry, e.g., problem-selection, the constitution of the scientific domain, the determination of facts, the very research results, and criteria of validity and truth. Science must be treated like any other symbolic form—namely as a mode of structuring reality embedded in the social structure of the whole society (p. 135)" (cf., Freedman, 1960) [4]. Lievrouw (1988, p.7*ff*) argues that researches are organized into four distinct "programs" of research:

1. Artifact studies: scientific information as an objective commodity, whose value is independent of its use;
2. User studies: scientific information as a commodity whose value depends on the practical needs of the user;
3. Network studies: scientific information as a social link, whose value is determined by its utility in the coherence of social networks;
4. Lab studies: scientific information as a social construction of scientists, with its value completely dependent on the changing perceptions of those individual scientists (so called because their authors typically employ participant observation or other ethnographic techniques to gather data in the scientists' workplace).

---

[4] See also Bernal, 1939; Bush, 1945; Callon, 1994; Etzkowitz and Leydesdorff, 1998; Johnson, 1972; Nelson, 1962; Nelson and Romer, 1996; Nordhaus, 1969; Rosenberg, 1974.

The evolution of scientific disciplines is critical to science and society to explain human progress. The most prevalent models of scientific development are:

- model of the accumulation of knowledge
- model of scientific paradigm shifts by Khun
- model of research programme by Lakatos
- approach by Tiryakian
- theoretical revisionism by Alexander Jeffrey
- model of openness, closure and branching described by Mulkay

The main characteristics of these approaches are briefly described as follows.

☐ The cumulative model of knowledge

The cumulative model states that scientific development is due to a gradual growth of knowledge based on a sum of facts accumulated by scholars (Haskins, 1965). In particular, science is an activity of accumulation (Science, 1965). Seidman (1987, pp. 121-122) argues that: "The cumulative addition of facts and verified propositions, conceptual refinements, or analytical developments dislodge erroneous theories, and propels us toward theories which are closer to the truth about society…. virtually every current social scientific theory strives to achieve legitimacy and dominance by reconstructing the past as a cumulative development crystallizing in its own systematization". The science evolves with a convergence among scientific fields that creates a deeper unity within the structure of science (Coccia and Wang, 2016; Haskins, 1965). Moreover, the evolution of science is irreversible and can never go back (Science, 1965).

☐ The model of scientific paradigm shifts by Khun

The scientific development is due to long periods of knowledge accumulation of "normal science"[5], interrupted by discontinuous transformations generated by new theoretical and empirical approaches that support the transition from an existing scientific paradigm to an emerging new paradigm. In fact, paradigm shifts are the major source of scientific change in society (Kuhn, 1962). Radical changes of theory can have a significant impact on several disciplines (e.g., the pervasive impact of artificial intelligence in different research fields; cf., Coccia, 2020) or these changes of theory can have consequences within a specific scientific discipline in which the change has taken place (e.g., the impact of the discovery of quasicrystals in to the field of condensed matter; cf., Andersen, 1998, p. 3; Coccia, 2016). Moreover, in this theory, scientific paradigm shift can be *major* in the presence of discontinuity with previous theoretical framework (e.g., target therapy vs. chemotherapy in cancer research; cf. Coccia, 2012b, 2012c, 2014a, 2015a, 2016a), and *minor* whether it generates continuity between successive paradigms (e.g., nanoparticle-delivered chemotherapy in oncology; Coccia and Wang, 2015). In general, major or minor paradigm shifts support the long-run evolution of science, disciplines and research fields over time.

☐ The model of scientific programme by Lakatos

Lakatos (1978) attempts to resolve the perceived conflict between Popper's falsificationism and the revolutionary structure of science described by Kuhn. Lakatos (1968, p. 168, original Italics and emphasis) argues that:

> science . . . can be regarded as a huge research program . . . .progressive and degenerating problem-shifts in series of successive theories. But in history of science we find a continuity which connects such series. . . . . The programme consists of methodological rules: some tell us what paths of research to avoid (*negative heuristic*), and others what paths to pursue (*positive heuristic*) - By 'path of research' I mean an objective concept describing something in the Platonic 'third world' of ideas: a series of successive theories, each one 'eliminating' its predecessors (in footnote 57) - . . . . What I have primarily in mind is not science as a whole, but rather particular research-programmes, such as the one known as 'Cartesian metaphysics. . . . a 'metaphysical' research-programme to look behind all phenomena (and theories) for explanations based on clockwork mechanisms (positive heuristic). . . . A research-programme is successful if in the process it leads to a progressive problem-shift; unsuccessful if it leads to a degenerating problem-shift . . . . Newton's gravitational theory was possibly the most successful research-programme ever (p. 169). . . . The reconstruction of scientific progress as proliferation of rival research-programmes and progressive and degenerative problem-shifts gives a picture of the scientific enterprise which is in many ways different from the picture provided by its reconstruction as a succession of bold theories and their dramatic overthrows (p. 182).

---

[5] " 'normal science' means research firmly based upon one or more past scientific achievements that some particular scientific community acknowledges for a time as supplying the foundation for its further practice" (Kuhn, 1962, p. 10, original emphasis).

Lakatos' model of the research programme is based on a hard core of theoretical assumptions that cannot be abandoned or altered without abandoning the programme altogether. The evolution of scientific fields here is due to the creation of a research programme that guides the subsequent scientific development of one or more research fields and/or disciplines over time (Lakatos, 1978). Finally, Lakatos' model provides for the possibility of a research programme that is not only continued in the presence of troublesome anomalies but that remains progressive despite them.

☐ The approach by Tiryakian for development of science

Tiryakian (1979) argues that the *scientific school* is the unit of analysis for a model of scientific development. Major schools guide the discipline by providing a new methodology or a new conceptual scheme of social reality. Tiryakian (1979) rejects both the empiricist approach that discoveries initiate scientific change and the rationalist claim that conceptual refinements of theoretical models stimulate a scientific change. In short, the formation of a school offers a new scientific direction to studying social reality that initiates significant scientific advances over time (e.g., in economics the Monetarism, started in 1950s with Milton Friedman, is a new school of thought based on control of money in the economy to affect price levels and economic growth *versus* Keynesian economics based on government expenditures with fiscal policy.

☐ The approach of revisionism by Alexander Jeffrey for scientific development

Unlike Kuhn (1962), Alexander (1979) proposes that scientific theories do not change in a revolutionary manner. Scientific theories are based on different autonomous entities, such as presuppositions, ideology, models, laws, concepts, propositions, methodology, etc. that shape science, articulate its problems, and have a distinctive mode of discourse with its own standards of assessment. Seidman (1987) argues that: "the discovery of anomalies or analytical criticisms of one or another dimension of a theory sets in motion a process of theoretical revision". In short, Tiryakian (1979) analyzes the tensions and dynamics of the social structure of the school and its relation to scientific community. By contrast, Alexander (1983, p. 349) argues that the engine of scientific change is due to new theoretical frameworks that generate a revision of current conceptual scheme, marking the life-history of a school.

☐ Models of scientific progress described by Mulkay (1975)

*The model of openness*

Scholars of the model of openness argue that science and technology is most likely to flourish in democratic society because science has democratic values and democratic nations do not have barriers on new results of scientific communities (cf., Coccia, 2010). In this context, discoveries and scientific breakthroughs can be advances of scientific knowledge if findings are made accessible to the critical inspection of other scholars in scientific community. In short, researchers have to communicate their new results and data to other scholars, facilitating reproducibility of results for validation of findings and/or new theories. Researchers, producing and sharing new breakthroughs and discoveries, are rewarded with a higher reputation and recognition in scientific communities that increases the traffic of their research articles and data, as well as it increases citations, funds for research, etc. (cf., Coccia, 2018c, 2019c). In fact, science, within open research communities and democratic settings, will grow rapidly because there is low resistance to new scientific ideas (De Solla Price, 1986; Kitcher, 2001; Merton, 1957; Mulkay, 1969; Coccia, 2010, 2017b). However, Max Planck (1950, pp. 33-34) states that: "a new scientific truth does not triumph by convincing its opponents and making them see the light, but rather because its opponents eventually die, and a new generation grows up that is familiar with it". For instance, the discovery of quasicrystals in 1982 by Shechtman et al. (1984) was a remarkable and controversial finding, violating the textbook principles of solid state materials. The interpretation that these materials represented a new type of solid was disputed vigorously, most notably by Pauling (1987), American Chemist with two Nobel Prizes. During the last decade of his life, Pauling tried to prove that quasicrystals are really just twinned periodic crystals. All his models were proven wrong. At the end of his life he remained the only prominent opponent to quasiperiodicity in crystals. Polanyi (1958,1963) argues that scientists are often not open-minded, independent puzzle-solvers, but rather men devoted to solving a limited range of problems rigidly defined by their scientific group. Hence, the evolution of science is due to: "a series of battles in which innovators have been forced to fight against the entrenched ideas of fellow scientists" (Mulkay, 1975, p. 12).

*The model of closure*

The history of science shows the existence of scientific orthodoxies, which tend to generate intellectual resistance in scientific progress (Cohen, 1952). This approach is consistent with the nature of scientific education that produces intellectual conformity from old generations of scholars to new ones. Mulkay (1975, p. 514) argues that the advances of scientific knowledge in Kuhn's theory are due to intellectual closure, rather than intellectual openness of scholars. The scientific evolution is due to an open rebellion against the existing paradigm created by intellectual orthodoxy (Cohen, 1952). In fact, scientific paradigm shift is mainly due to an accumulation of anomalies that cannot be answered within existing scientific rules or theories. These anomalies or limitations of existing paradigms lead to few scholars to think in wholly new directions, changing accepted paradigms in science and giving a new conceptual scheme (Boring, 1927). For instance, Büttner et al. (2003, pp. 38-39) state that in 1900, the establishment of the radiation spectrum by precision measurements and its description by Planck's formula creates an anomaly and a crisis in classical physics. Max Planck attempts to derive his radiation formula on the basis of classical physics, involving in an error. Einstein discovers the error in Planck's classical derivation and lays to the establishment of a quantum derivation of the radiation law. This crisis discards an existing scientific paradigm and establishing aspects of an emerging new paradigm that, however, was not immediately recognized as the solution of the problem. The authoritative lecture in 1908 by the recognized master of classical physics, H.A. Lorentz, validated the discovery and the widespread acceptance of the new paradigm in physics. Another driver of scientific development is new technologies that destroy existing paradigms creating new theoretical frameworks, such as transmission electron microscopy[6] and associated high-energy electron diffraction[7] have supported the discovery in 1982 of quasicrystals by Daniel J. Shechtman that investigated rapidly-quenched phases in alloys of Aluminum at the National Bureau of Standards, USA (Shechtman et al. 1984; cf., Coccia, 2016, 2019d). This discovery, according to Thiel (2004, p. 69), suggests that: "solids can adopt structures that are atomically well-ordered (giving rise to discrete diffraction patterns), yet not periodic (since *n*-fold rotational axes cannot exist in a conventional periodic material unless $n = 2$; 3; 4, or 6". Levine and Steinhardt (1984) claim that this breakthrough lays the foundations for the concept of 'quasicrystallinity': a new type of organization in the condensed matter. Moreover, this discovery violates the principles of solid state materials and the definition of crystals by IUCr-The International Union of Crystallography (1992)[8], prior 1992, such that this International Scientific Union adhering to the International Science Council provided a new definition of crystal that validates the scientific paradigm shift in crystallography based on discovery of quasi-periodic crystals.

*The model of branching in science*

Science can evolve with social and research networks of scholars (Adams, 2012, 2013). In fact, Adams (2012, p. 335) claims that: "New collaboration patterns are changing the global balance of science". The evolution of any one research network depends considerably on developments in neighboring scientific fields in the geography of science. Mulkay (1975) argues that the exploration of a new research field is usually set in motion by a process of scientific migration of scholars in the presence of specific characteristics, such as established research networks are declining in terms of significant results (Mullins, 1973; Coccia, 2018). In this model, leading scholars, starting from new scientific breakthroughs, create research teams and international scientific collaborations that lay the foundations for developing new research fields (cf., Coccia, 2018). For instance, Relman (2002), American microbiologist, produces one of the first articles that investigates the human microbiome, creating a research team−The Relman Lab−within Stanford University School of Medicine and VA Palo Alto Health Care System in California to develop the general themes of host-microbe interactions and human microbial ecology (Coccia, 2018). This scientific breakthrough has created a new research field, based on a broad cross-section of sub-disciplines within microbial ecology, in which many scholars collaborate, spreading their results on new journals, which bring together scientific communities working in the environmental, animal and biomedical microbiome arenas, for presenting new researches and methodologies, as well as for discussing current and future trends in microbiome research.

---

[6] Transmission Electron Microscopy (TEM) is a radical innovation that operates with electrons that are accelerated to a velocity approaching the speed of light (Coccia, 2016); the associated wavelength is five orders of magnitude smaller than light wavelength and the resolution of the material imaging and structure determination is at atomic level (Hawkes, 2007; Fultz and Howe, 2007; Reimer and Kohl, 2008). TEM is a microscopy technique that can provide information of the surface features, shape and structure and is an appropriate instrument to support scientific advances in cancer research, materials science, semiconductor research, metallurgy, and so on (Coccia, 2012, 2016).
[7] High-energy electron diffraction is a technique used to characterize the surface of crystalline materials.
[8] The International Union of Crystallography, prior to 1992, defined the crystal: "a substance in which the constituent atoms, molecules or ions are packed in a regularly ordered, repeating three-dimensional pattern. Among the rotational symmetries two-, three-, four- and six-fold axes are allowed, while five-, seven- and all higher rotations are disallowed".

In this context, Sun et al. (2013, p. 4) claim that the socio-cognitive interactions of scientists and scientific communities play a vital role in shaping the evolution of scientific fields. Sun et al. (2013) also argue that research fields evolve from diversification and/or merger of scientific communities within collaboration networks. This literature of social construction of science has investigated international collaborations between research organizations because foster scientific breakthroughs, technological advances, and other events that are fundamental determinants of the social dynamics of science[9]. In fact, Morillo et al. (2003, p. 1237) claim that research fields are increasing the interdisciplinary because of a combination of different bodies of knowledge and new communities of scholars from different disciplines that endeavor to solve more and more complex problems in nature and society[10]. Another distinct class of approaches analyzes the patterns of basic and applied sciences (Boyack, 2004; Boyack et al., 2005; Frame and Carpenter, 1979; Klavans and Boyack, 2009; Simonton, 2004; Smith et al., 2000). In fact, social studies of science argue that basic research is aiming at finding truth, whereas applied research is aiming at solving practical problems (Kitcher, 2001; Frame and Carpenter, 1979; Fanelli and Glänzel, 2013). Frame and Carpenter (1979) suggest that basic fields include mathematics, astronomy (similar to space science), physics and chemistry; and applied research fields include biology, clinical medicine, and engineering/technology. Storer (1967) focuses on the concept of hard and soft to characterize different branches of science. In particular, Storer (1967, p. 75, original emphasis) claims that: "The degree of rigor seems directly related to the extent to which mathematics is used in a science, and it is this that makes a science 'hard' "; this approach suggests that chemistry and physics have about the same "rated hardness" i.e., they are characterized by a high degree of rigor. Nevertheless, these research topics are the subject of ongoing discussion because scientific fields are dynamic entities that evolve over time with a pattern of convergence between basic and applied sciences (Coccia and Wang, 2016; cf., Sintonen, 1990).

---

[9] cf., Beaver and Rosen, 1978; Coccia and Bozeman, 2016; Coccia and Wang, 2016; Coccia and Rolfo, 2009; Coccia et al., 2015; De Solla Price, 1986; Frame and Carpenter, 1979; Latour, 1987; Latour and Woolgar, 1979; Mulkay, 1975; Newman, 2001; Sun et al., 2013; Storer, 1970.

[10] Coccia, 2012, 2012a; Fanelli and Glänzel, 2013; Gibbons et al., 1994; Guimera et al., 2005; Kitcher, 2001; Klein, 1996; Sun et al., 2013; Wagner, 2008.

One stand of this literature emphasizes the empirical properties of the evolution of science. Coccia (2018), analyzing the emerging research fields of human microbiome, evolutionary robotics and astrobiology (also called exobiology), suggests some properties of the evolution of research fields, such as: 1) the evolution of a research field is driven by few disciplines that generate more than 80% of documents (concentration of scientific production); 2) the evolution of research fields is path-dependent of critical disciplines: they can be parent disciplines *or* new disciplines emerged during the evolution of science from a process of convergence of different research fields; 3) in particular, the evolution of research fields can be also due to new disciplines originated from a process of specialization within applied or basic sciences and/or a convergence between disciplines. Finally, 4) the evolution of research fields can be due to both applied and basic sciences. In general, these studies show that scientific fields are not static entities but they change with the evolution of science and society (Coccia and Wang, 2016; Coccia, 2018; Sun et al., 2013). Some of these changes are progressive processes because of the essential nature of scientific progress in society (Simonton, 2004, p. 65). Overall, then, while several studies exist in social studies of science, scientometrics and sociology of knowledge, many general characteristics and properties of endogenous processes underlying the evolution of research fields in applied sciences are still unknown. This paper here endeavors to analyze the evolution of experimental physics to suggest general properties of the dynamics of applied sciences.

**Materials and Methods**

The concept of discipline in science derives from Latin *disciplina,* derivation of *discĕre*= to learn. In particular, scientific discipline is a system of organized and systematized norms, theories and principles, established and developed by specific methods of inquiry (Coccia and Benati, 2018). A research field is a sub-set of a discipline that investigates specific research topics to solve theoretical and practical problems that can generate science advances of applied and/or basic sciences in society[11]. Experimental physics is a vital applied science in physics. In fact, progress in science is made possible by a comparison of the measured behavior of real world with expectations from theory. The experimental physics is a discipline where physics scholars have an intensive laboratory experience that concentrates on experiments for substantiating and/or challenging established and/or new theories in physics. The disciplines and sub-fields of research in experimental physics concern with the observation of physical phenomena and experiments[12]. In particular, experimental physics regroups all the research fields of physics that focus on data acquisition, data–acquisition methods, and the detailed conceptualization and realization of laboratory experiments. It is often put in contrast with theoretical physics, which predicts and explains the physical behavior of nature, rather than acquire data and provide evidence about it.

- *Data and their sources*

Data under study here are more than 121,500 document results in experimental physics (at October 2019). The source of these data is ScienceDirect (2019) and its tool of Advanced Search to find scientific products that have in title, abstract or keyword the following term: "experimental physics". This study focuses on followings information downloaded: scientific products per year, keywords, year of the first scientific product. Moreover, keywords detected in experimental physics provide main information about research fields, sub-domains of research and new technologies supporting the evolution of this and other disciplines in physics over the course of time.

---

[11] This study uses the terms of research field, research topic or keyword within scientific products like interchangeable concepts because the difference between these different types is difficult to identify in science domains.
[12] Main research topics in experimental physics are described by: Barger and Olsson, 1973; Bleaney and Bleaney, 1965; Cheng, 2010; Halliday et al., 2014; Heyde, 1994; Jackson, 1999; Kleppner and Kolenkow, 2014; Lilley, 2001; Martin, 2006, Martin and Shaw, 2008; Perkins, 2000; Phillips, 1994; Squires, 2001; Taylor, 1997; Young and Freedman, 2012.

- *Measures*

– Research fields and sub-fields of research underlying experimental physics are detected with keywords provided by the tool of "Advanced Search" in ScienceDirect (2019), using the term of "experimental physics". After that, each keyword is further investigated using appropriate filter that provides scientific products per years.

– The scientific aspects, underlying experimental physics, are measured considering the type of research field (detected with keyword) and number of articles and all scientific products (articles, conference papers, conference reviews, book chapters, short surveys, letters, etc.).

– The temporal aspects of the evolution of research fields are measured with the period starting from the first year of scientific products including the keyword to 2019.

The evolution of sub-fields of research within experimental physics, measured with the number of articles and other scientific products, is important for understanding characteristics of the dynamics of this scientific discipline and in general of applied sciences.

- *Data analysis procedure*

– <u>Comparative study of old and new syllabi in experimental physics</u>

Firstly, this study compares the content of the book of experimental physics by Genovesi (1786) *Elementi di Fisica Sperimentale*, Napoli (Italy), one of the first books of experimental physics for academic institutions, with some modern syllabi of experimental physics in Europe and USA about 2010s. This comparative analysis shows, *ictu oculi*, the evolution of this discipline over a period of about 250 years, considering what research topics decline, what research topics emerge, evolve, transform and/or grow considerably to create new research fields that evolve as autonomous entities.

- <u>Chronologies, weight of sub-domains of research and historical period of new research fields in experimental physics</u>

This study presents bar graphs of the first 40 keywords in experimental physics with the highest number of scientific products and the year of the first appearance of papers studying this research topic, sub-field of research or new technology (*keyword can identify different scientific subjects, i.e., research field, research topic or new technology*). This empirical analysis shows the role of new and old research fields in the evolution of experimental physics also in terms of total number of scientific products. The study also considers the chronology of research fields within experimental physics, showing the timeline of these research fields based on the first year of occurrence of the research topic in experimental physics that in 2019 has a high number of scientific products. Moreover, this study divides keywords within experimental physics in two sets:

- Research fields originated *pre*-1900s
- Research fields originated *post*-1900s

The analysis provides information to calculate the average timing in years of occurrence of new research fields in experimental physics *pre-* and *post*-1900, as well as how many new research fields or new research topics emerge in average every decade in experimental physics (*pre-* and *post-* 1900).

The Weight of Research Fields (WRFs) is given by:

$$WRF = \frac{\text{number of scientific products in a research field or specific set of research fields}}{\text{total number of scientific products of all research fields (or a wider set)}}$$

In particular, the analysis calculates the ratio WRFs considering research fields originated *post*-1900 divided by a wide set of 40 research fields with the highest number of scientific products. In this context, it is calculated the historical period in years of research field=2019-*y1*, where *y1* is the year of the first paper using the research topic/keyword under study. This empirical analysis shows the average age of research fields and topics in experimental physics originated *post*-1900 and also the average age of the newest research fields originated during the 1940s.

- <u>Upwave of scientific cycle of new research fields in experimental physics</u>

The evolution of research fields in experimental physics is also investigated considering the upwave of scientific cycle given by: AM ($i$)= length in years of the upwave of scientific cycle of research field $i$

AM ($i$)= M$i$ −A$i$

A$i$= year of the first paper including the term about the research field $i$ in experimental physics

M$i$ = year of the peak of scientific production of the research field $i$ in experimental physics

After that, the arithmetic mean is calculated to detect the average cycle of upwave of all research fields in experimental physics, as follows:

$$Average\ period\ \overline{AM}\ of\ research\ fields\ in\ experimental\ physics$$

$$= \sum_{i=1}^{N}\left(\frac{M_i - A_i}{N}\right) \quad \text{with } i=1, 2, \ldots, N \text{ (research fields)}$$

- <u>Rate of evolutionary growth of research fields and scientific forecasting in experimental physics</u>

The analysis here also provides trends of research fields in experimental physics originated *post*-1900 to assess which research fields are likeliest to evolve rapidly in this applied science.

The preliminary statistical analysis is also based on descriptive statistics of research fields in experimental physics: arithmetic mean (M), standard deviation (SD), skewness and kurtosis of scientific products over time. This preliminary analysis is important to verify the normality of distribution and apply appropriate parametric analyses.

The main statistical analysis focuses on the evolution of research fields, originated *post*-1900, with the highest number of scientific production, considering the scientific production as a function of time on a semi-logarithmic paper.

Model is specified as follows:

$$\ln Y_i = \beta_0 + \beta_1 \, t + \varepsilon_t \qquad [1]$$

$Y_i$= scientific production in the research field *i (i=1, …, N)*
$\beta_0$ is a constant
$\beta_1$ is the coefficient of regression
*t* is time
$\varepsilon_t$ is error term

The coefficient of regression provides a preliminary assessment of the rate of evolution of research fields underlying experimental physics. This model also generates predicted values. Finally, the scientific forecasting of new research fields in experimental physics is performed as follows: the procedure in Statistics Software SPSS selects *Time* as independent variable, whereas dependent or response variable is scientific production of research fields; after that we use all cases to predict values using the prediction from estimation period through last case in the SPSS Statistics Software. Results of scientific forecasting are represented with sequence chart, using the natural logarithm of predicted values from linear model of scientific production of research fields as function of time [1]. This analysis can show future driving research fields in experimental physics.

These relationships [1] for empirical analysis and scientific forecasting are investigated using ordinary least squares (OLS) method for estimating the unknown parameters in a linear regression model. Statistical analyses are performed with the Statistics Software SPSS® version 24.

**Results**

☐ *Comparatives analysis of old and new syllabi in experimental physics*

The experimental physics is a critical research field in applied sciences and in order to analyze its evolution, this study shows the content of the book in experimental physics by Antonio Genovesi (1786) *Elementi di Fisica Sperimentale*, Napoli (Italy), one of the first books for higher education in Europe, originally written in Latin. The content of this book is compared with new syllabi of experimental physics in Europe and the USA in 2010s to see what research topics of experimental physics−over a period of 250years−are still present, what are declined, what research fields are emerged, evolved, transformed and/or grown considerably to create new research fields in physics. The comparative contents of syllabi in Table 1 show, *ictu oculi*, how in 1780s experimental physics was a wide discipline, including not only topics of physics still taught today adding, of course, new theories, but it also included topics of astronomy, physical geography, geology, zoology, seismology, medicine and botany. This comparison suggests that the evolution of experimental physics has generated a *scientific fission*: the division of a research field over time into more research fields that evolve as autonomous entities. In particular, the scientific fission of experimental physics has produced multiple research fields that today evolve autonomously in science, such as astronomy, physical geography, geology, seismology, etc. These research fields, in turn, during the evolution of science assume the status of scientific disciplines that generate further scientific fissions, also driven by new technologies, creating new research fields, such as astronomy has generated radio-astronomy, cosmic rays, astrophysical fluids and plasmas, extragalactic astronomy and cosmology, interstellar medium and star formation, stellar astrophysics, exoplanet systems, etc.[13]

---

[13] In 2010s, other research fields of physics are: Galaxy Formation, Particle Physics, Early Universe, Quantum Mechanics, Quantum Physics and Relativity, Random Processes in Physics, Solid State Physics, Atomic Physics, Galaxies, Photonics, Exoplanets, Nuclear Fusion and Astrophysical Plasmas, Superconductors and Superfluids, Quantum Field Theory, Radio Astronomy, Photon Science, Gauge Theories, Stars and Stellar Evolution, Soft Matter Physics, etc.

**Table 1** – Comparison of the content of syllabi in experimental physics (1780s vs. 2010s)

| Genovesi (1786) *Elementi di Fisica Sperimentale* Title in English is: *Fundamentals of Experimental Physics* | Examples of course syllabi in experimental physics at Polytechnic school of engineering in Europe and U.S. Universities (2010s) |
|---|---|
| BOOK 1<br>Nature of physics, principles and elements of the universe<br>Rules of philosophical speculation in physics<br>Universe<br>Principles and elements<br>Greek philosophy, modern philosophy: Galileo and Descartes<br>Newton's philosophy, philosophy of Leibniz and Wolff<br>General properties of bodies<br>Vacuum<br>Divisibility of bodies<br>Gravitation<br>Rules of time and motion, general rules of motion, compound motion<br>Force and power<br>Resistance and oscillation of pendulum<br>Attraction and attraction of the magnet<br>Attraction of fluids and repulsion<br>Electricity<br>Particular properties of bodies<br>Fluidity in general<br>Fluid action<br>Hardness, fragility, softness, flexibility and elasticity<br>Opaque, diaphanous and luminous body<br>Reflection and refraction of light<br>Eye structure<br>Opacity and colors<br>Fire, heat, cold, thermoscopes and thermometers<br>Sound<br>Smell and taste<br>BOOK 2<br>Artificial sphere called armillary<br>Celestial poles, axis of the Earth, equators, parallels, and circles<br>Horizon, polar regions, derivative circle, meridian, triple position of the sphere, height<br>Sun, moon and other planets, comets and stars<br>World system<br>Criticism of the Copernican system<br>Causes of celestial motions<br>Earth and sea system<br>Theory of the interior of the Earth<br>Internal bodies of the Earth: sulfur and bitumen<br>Earthquakes<br>Metals, fossils<br>Waters, sources, rivers and the nature of the sea<br>Animals and plants<br>Structure of the human body<br>Circulatory system and heart<br>Glands of the organism<br>Digestive system<br>Feeding and breathing of animals<br>Movement of animals and muscles<br>Brain and the nervous system<br>Perfect and imperfect animals<br>Plants and diffusion<br>Air and meteors<br>Meteors, colored and non-colored waters<br>Igneous meteors<br>Wind | Europe (cf., Politecnico di Milano, 2019)<br><br>MECHANICS<br>Kinematics and dynamics of the point<br>Work and energy<br>Gravitational field<br>Elements of the dynamics of points and rigid bodies<br><br>THERMODYNAMICS<br>Temperature, heat and work<br>Principles of thermodynamics<br><br>ELECTROSTATIC AND MAGNETOSTATIC<br>Field and electrostatic potential<br>Conductors and dielectrics<br>Electric current in the conductors<br>Magnetic field and magnetic field sources<br>Phenomenology of magnetic materials<br><br>In addition, the courses include activities in laboratory.<br><br>U.S.A.<br>The course concentrates on experiments in 20th-century physics, e.g. the quantum nature of charge and energy, etc.<br><br>In particular, some experiments held in university courses are (cf., NYU Department of Physics, 2019):<br><br>☐ The Hydrogen-Deuterium Isotope Shift<br>☐ Relativistic Electron Momentum<br>☐ The Muon Lifetime<br>☐ Pulsed Magnetic Resonance and Spin Echo<br>☐ Rutherford Scattering<br>☐ The Mossbauer Effect<br>☐ Magnetic Susceptibility Under Phase Transitions<br>☐ Optical Pumping of Rubidium<br>☐ Diode Laser Spectroscopy<br>☐ Laser Particle Trapping and Brownian Motion<br>☐ Quantized Conductance<br>☐ Quantum Optics of Photon Pairs |

Figure 1 shows the first 40 keywords in experimental physics that have the highest number of scientific documents.

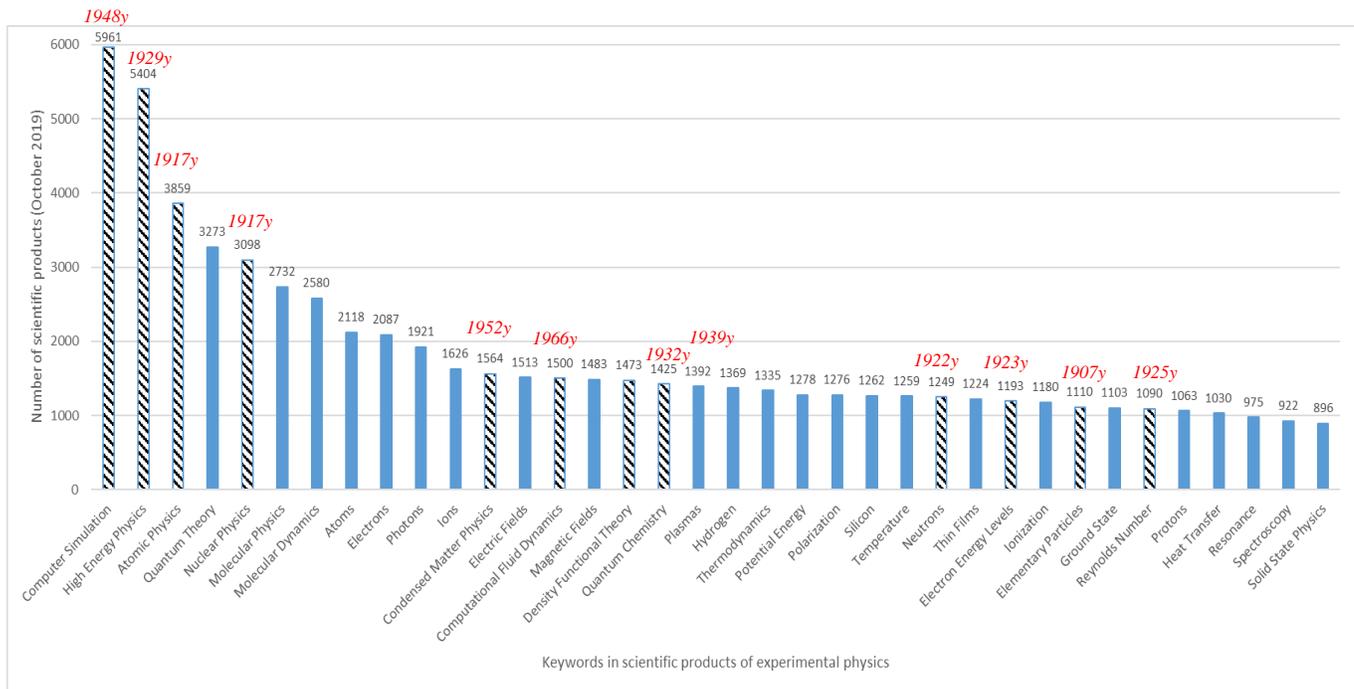

**Figure 1.** Keywords in experimental physics with the highest number of scientific products
*Note*: Research fields originated *post*-1900 are bars with diagonal stripes, the year (*y*) of the first paper having this keyword is in the top of the bar (in red and Italics). In the top of the bar, below the year, the number indicates the total scientific products as detected by ScienceDirect (2019).

Figure 1 shows that 4 of the first 5 research fields/keywords supporting the evolution of experimental physics are emerged *post*-1900s:

− n.1, computer simulation with 5,961 scientific products; the first paper using this keyword is in 1948

− n.2, high energy physics with 5,404 scientific products; the first paper using this keyword is in 1929

− n.3, atomic physics with 3,859 scientific products; the first paper using this keyword is in 1917

− n.5, nuclear physics with 3098 scientific products; the first paper using this keyword is in 1917

This result suggests that the evolution of research fields is driven mainly by new research topics. Total number of scientific products of the first 40 research fields is 69,179 documents (cf. Tab. 2 and Fig. 2). Moreover, 15 research fields on 40 ones with the highest number of scientific products within experimental physics, have emerged *post*-1900, i.e., about 38%, whereas 62% are started *pre*-1900.

**Table 2** – Scientific weight (WRF) of the first 15 research fields in experimental physics *post*-1900

| N | Research fields with the highest number of scientific products in experimental physics, originated *post*-1900 | Year of the first paper using the keyword (in decreasing order) $y1$ | Number of scientific products with this keyword | Age (in *years*) of research field $=2019-y1$ |
|---|---|---|---|---|
| 1 | Computational Fluid Dynamics | 1966 | 1,500 | 53 |
| 2 | Tokamak Devices | 1965 | 860 | 54 |
| 3 | Condensed Matter Physics | 1952 | 1,564 | 67 |
| 4 | Computer Simulation | 1948 | 5,961 | 71 |
| 5 | Density Functional Theory | 1939 | 1,473 | 80 |
| 6 | Solid State Physics | 1936 | 896 | 83 |
| 7 | Quantum Chemistry | 1932 | 1,425 | 87 |
| 8 | High Energy Physics | 1929 | 5,404 | 90 |
| 9 | Reynolds Number | 1925 | 1,090 | 94 |
| 10 | Electron Energy Levels | 1923 | 1,193 | 96 |
| 11 | Neutrons | 1922 | 1,249 | 97 |
| 12 | Anisotropy | 1922 | 835 | 97 |
| 13 | Atomic Physics | 1917 | 3,859 | 102 |
| 14 | Nuclear Physics | 1917 | 3,098 | 102 |
| 15 | Elementary Particles | 1907 | 1,110 | 112 |
| | Total scientific products of 15 research fields | | 31,517 (A) | $M_{1-15}=85.67y$ |
| | Total scientific products of the first 40 research fields | | 69,179 (B) | $SD_{1-15}=17.63y$ |
| | Weight of new research fields on 40 ones (A/B)×100 | | 46% | $M_{1-5}=65.0y$ |
| | All scientific products in experimental physics (11 October 2019) | | 121,722 (C) | $SD_{1-5}=11.5y$ |
| | Weight of 15 research fields on all research fields (A/C)×100 | | 26% | |

*Note*: M=Arithmetic Mean, SD=Standard Deviation.

The 15 research fields emerged *post*-1900 have a total number of scientific products equal to 31,517 (46% on a total of the first 40 research fields), whereas the other research fields *pre*-1900 have, of course, a percent weight equal to 54% on a total of 40 research fields. This result confirms that the driving forces fields in applied sciences seem to be due to new research fields, rather than older ones. Moreover, table 2 shows that these research fields have an average age of about 85 years (SD=17.6*y*); the newest research fields originated during the 1940s are five with an average age of 65 years and driven mainly by rapid development of computer technologies, such as computational fluid dynamics, tokamak devices, condensed matter physics, computer simulation in physics, density functional theory, etc. (cf., Fig. 2). As a matter of fact, researchers in physics develop simulation methods based on statistical mechanics/quantum mechanics, numerical analysis and data structures to investigate and solve more and more complex problems and produce quantitative predictions in manifold branches of physics. For instance, density functional theory has made it possible for quantum chemistry calculations to reach accuracies comparable to those obtained in experiments for molecules of moderate sizes.

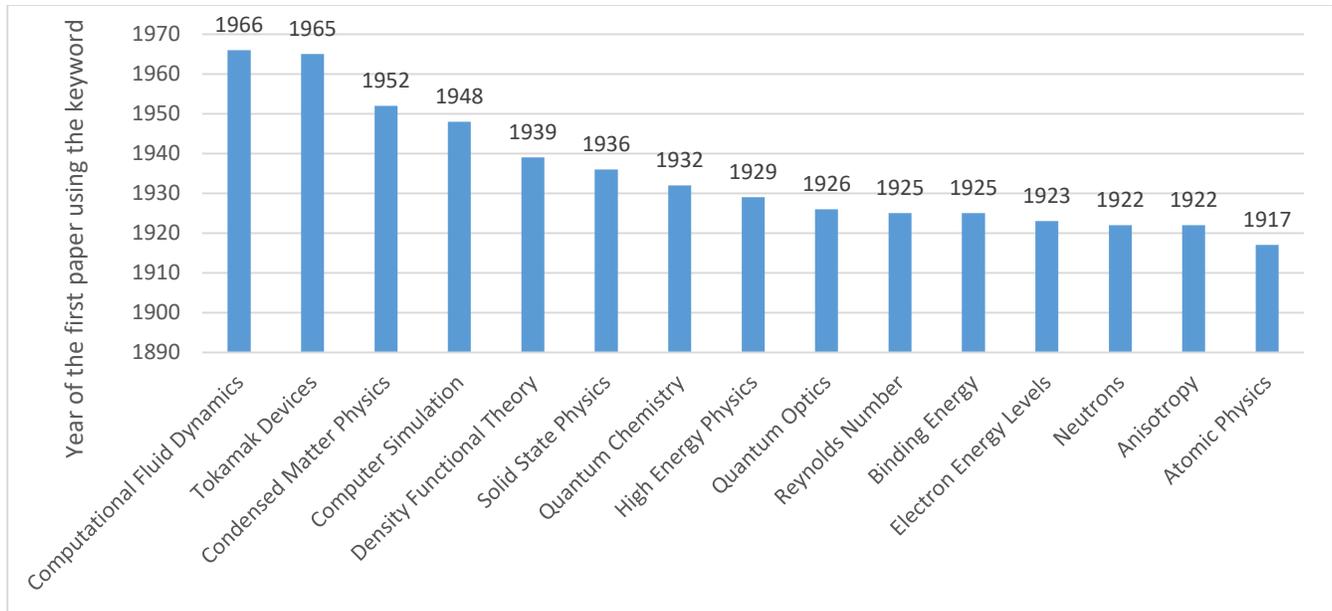

**Figure 2.**  The first 15 research fields emerged post-1900 with the highest scientific production in experimental physics; the order on *x*-axis is from the newest to oldest (from left to right)

In addition, analyzing 15 research topics started *post*-1900 that have the highest scientific production in experimental physics versus the other 25 research fields emerged *pre*-1900, findings in table 3 suggest that after 1900, new research topics emerge in average every 3.9 years (SD=3.6*y*), or every decade it emerges in average 2.5 new research topics (SD=1.5). The period pre-1900, considering research fields under study with the highest number of scientific products, shows that new research fields emerge in average every 2.3 years (SD=3.04*y*), as well as about 4.2 new concepts every decade (in average, SD=3.5).

**Table 3** – Timing of the emergence of new research fields in experimental physics (in years) and number of new concepts *pre-* and *post-*1900 in every decade

| Research fields $i$ *post-* 1900 | year of initial studies in increasing order | Difference in years $(t+1)_i - t_i$ | Number of new concepts every decade | Research fields $i$ *pre-* 1900 | year of initial studies in increasing order | Difference in years $(t+1)_i - t_i$ | Number of new concepts every decade |
|---|---|---|---|---|---|---|---|
| Shear Flow | 1911 | | | Resonance | 1843 | | 1 |
| Atomic Physics | 1917 | 6 | | Thermal Effects | 1853 | 10 | |
| Nuclear Physics | 1917 | 0 | 3 | Hydrogen | 1857 | 4 | |
| High Energy Physics | 1923 | 6 | | Oxygen | 1858 | 1 | |
| Reynolds Number | 1925 | 2 | | Polarization | 1859 | 1 | 4 |
| Binding Energy | 1925 | 0 | | Temperature | 1861 | 2 | |
| Quantum Optics | 1926 | 1 | | Ions | 1867 | 6 | |
| High Energy Physics | 1929 | 3 | 5 | Spectroscopy | 1869 | 2 | 3 |
| Quantum Chemistry | 1932 | 3 | | Quantum Theory | 1870 | 1 | |
| Solid State Physics | 1936 | 4 | | Molecular Physics | 1870 | 0 | |
| Density Functional Theory | 1939 | 3 | 3 | Electrons | 1870 | 0 | |
| Computer Simulation | 1948 | 9 | 1 | Photons | 1870 | 0 | |
| Condensed Matter Physics | 1952 | 4 | 1 | Electric Fields | 1870 | 0 | |
| Tokamak Devices | 1965 | 13 | 2 | Magnetic Fields | 1870 | 0 | |
| Computational Fluid Dynamics | 1966 | 1 | | Thin Films | 1870 | 0 | |
| | | | | Heat Transfer | 1870 | 0 | |
| | | | | Thermodynamics | 1871 | 1 | |
| | | | | Potential Energy | 1872 | 1 | 11 |
| | | | | Ground State | 1872 | 0 | |
| | | | | Plasmas | 1881 | 9 | |
| | | | | Fluid Dynamics | 1883 | 2 | |
| | | | | Molecular Dynamics | 1885 | 2 | 4 |
| | | | | Ionization | 1885 | 0 | |
| | | | | Astrophysics | 1892 | 7 | |
| | | | | Protons | 1898 | 6 | 2 |
| *Arithmetic mean (M)* | | *3.9 years* | 2.5 | | M | 2.29 years | 4.20 |
| *Standard Deviation (SD)* | | *3.6 years* | 1.5 | | SD | 3.04 years | 3.54 |

*Note*: alternate grey and white horizontal areas indicate a decade.

Moreover, results show that a cluster of new research fields/topics is emerged between World War I (WWI) and WWII and during the 1870s. Coccia (2018a) shows that structural changes of warfare can generate huge demand-side effects and powerful supply-side effects to support the evolution of science and technologies, clusters of innovation, new discoveries and other scientific/technological advances. In particular, the analysis here seems to reveal general sources of the evolution of scientific and technological change, rooted-in-war, that generates economic and social change (cf., Coccia, 2015, 2017).

☐ *Upwave of scientific cycle in experimental physics*

Another interesting results of this study is the duration of upwave of scientific cycle given by the difference between the year of emergence of the first scientific product in a specific research topic, until the year of peak of scientific production in the research field under study (Table 4). For the sake of briefness, this study considers some key research fields. Of course, this study does not consider research fields that are still growing over time, because we do not know the year of the future peak of scientific production. Results suggest that average period of the upwave of scientific cycle is about 80 years (SD is roughly 13 years).

Table 4 – Average duration of the upwave of scientific cycle in some research fields of experimental physics

| Research Field $i$ | Starting year of the first paper ($A_i$) | Year of the peak of scientific production ($M_i$) | Upwave of scientific cycle (in years), $AM_i = M_i - A_i$ |
|---|---|---|---|
| Condensed Matter Physics | 1952 | 2012 | 61 |
| Quantum Chemistry | 1932 | 2015 | 84 |
| High Energy Physics | 1929 | 2012 | 84 |
| Atomic Physics | 1917 | 2008 | 92 |
| *Arithmetic mean M (years)* | | | *M=80.25y* |
| *SD (years)* | | | *SD=13.38y* |

*Note*: M=arithmetic mean; SD=Standard Deviation.

❑ *Rate of evolutionary growth of research fields, drivers and scientific forecasting in experimental physics*

This study also analyzes the evolutionary growth of emerging research fields applied in experimental physics (Figg. 3-4). First of all, descriptive statistics shows the normality of distribution of variables under study based on coefficients of skewness and kurtosis in order to apply appropriate parametric analyses, using the growth model with equation $lnY = b_0 + b_1 \cdot t$. The relationships are investigated using ordinary least squares (OLS) method for estimating the unknown parameters in a linear regression model.

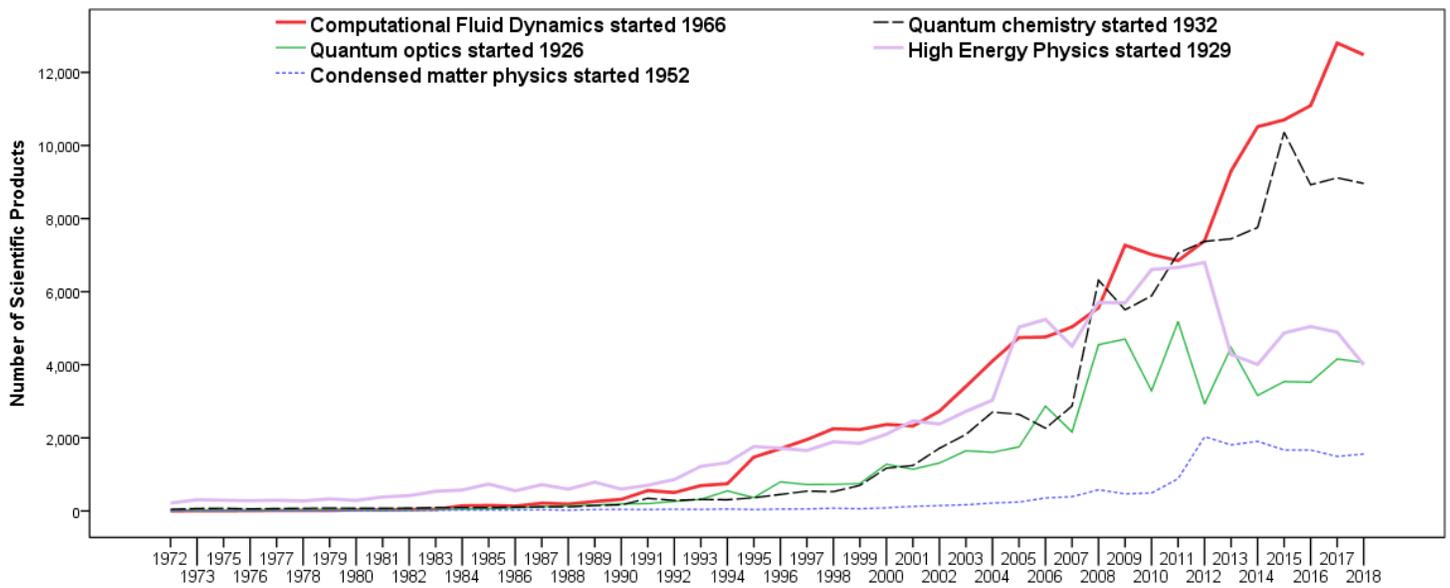

**Figure 3.** Trends of the scientific production of emerging research fields started *post*-1900 in experimental physics

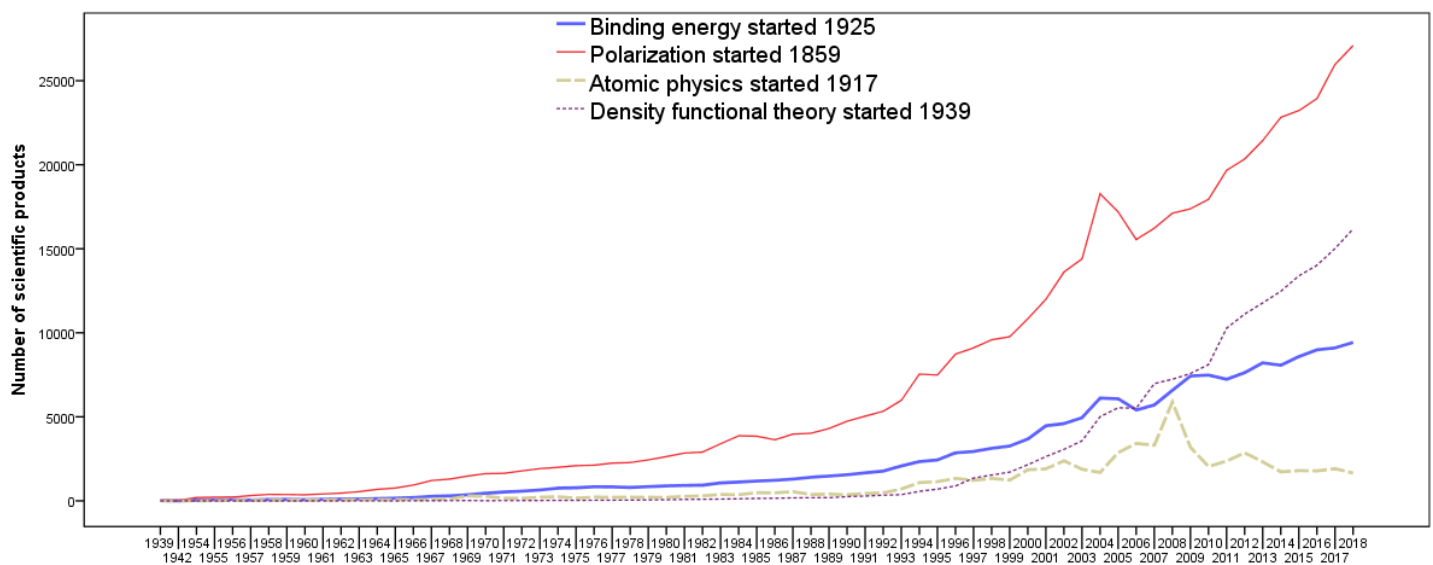

**Figure 4.** Other trends of the scientific production of emerging research fields started *post*-1900 in experimental physics

Statistical analyses are in table 5. In particular, emerging research fields in experimental physics with a high rate of growth are given by Computational Fluid Dynamics started in 1966 (b=0.19, *p-value*<.001), Density Functional Theory started in 1939 (b=0.15, *p-value*<.001), Condensed Matter Physics started in 1952 (b=0.14, *p-value*<.001). Findings also show that older research fields in experimental physics have a lower rate of growth, such as Polarization (started in 1859) has b=0.07 (*p-value*<.001), Atomic Physics (started in 1917) has b=0.09 (*p-value*<.001), etc.

**Table 5** – Estimated relationships of scientific production as a function of time (growth model) of emerging research fields in experimental physics

*Dependent variable*: *ln* scientific products concerning emerging scientific fields in experimental physics

| Research fields | Constant $b_0$ (St. Err.) | Coefficient $b_1$ (St. Err.) | F | $R^2$ adj. S=St. Err. of the Estimate |
|---|---|---|---|---|
| Computational Fluid Dynamics N= 52 (1966-2018) | −367.75*** (12.04) | 0.19*** (0.006) | 964.58*** | 0.95 S=0.61 |
| Condensed Matter Physics N= 66 (1952-2018) | −264.58*** (7.77) | 0.14*** (0.004) | 1192.38*** | 0.96 S=0.48 |
| Density Functional theory N= 82 (1939- 2018) | −301.29*** (5.65) | 0.15*** (0.003) | 2938.28*** | 0.98 S=0.47 |
| Quantum Chemistry N= 86 (1932-2018) | −233.23*** (5.54) | 0.12*** (0.003) | 1846.99*** | 0.96 S=0.54 |
| High Energy Physics N= 89 (1929-2018) | −203.32*** (5.51) | 0.11*** (0.003) | 1430.44*** | 0.95 S=0.63 |
| Quantum Optics N= 92 (1926-2018) | −240.27*** (8.43) | 0.12*** (0.004) | 843.45*** | 0.93 S=0.76 |
| Binding Energy N=93 (1925-2018) | −192.69** (4.77) | 0.10*** (0.002) | 1724.90*** | 0.95 S=0.63 |
| Atomic Physics N=101 (1917-2018) | −179.02*** (4.65) | 0.09*** (0.002) | 1558.03*** | 0.94 S=0.65 |
| Polarization N= 159 (1859-2018) | −126.90*** (2.99) | 0.07*** (0.002) | 1935.36*** | 0.93 S=0.89 |

*Note*: Explanatory variable is *time in years*. N is the number of observations from the specified time period (the first year indicates the first utilization of the concept in a scientific product of experimental physics, the second year is 2018 because 2019 is still ongoing). *** significant at 1‰; the standard errors of the constant and regression coefficient are given below in parentheses. *F* is the ratio of the variance explained by the model to the unexplained variance; $R^2$ adjusted is the coefficient of determination adj., below *S* is the standard error of the estimate.

In general, underlying research fields supporting critical disciplines tend to have a life cycle and maturity phase similar to other phenomena studied in biological and social sciences (Haskins, 1965); in particular, new research fields have a higher rate of growth than old ones, assuming the characteristic of driving forces in experimental physics.

**Table 6** – Drivers of experimental physics, considering the highest occurrence of keywords in scientific production (in each column items are in decreasing order of scientific production)

| Research fields | Sub-domains of research | New technologies / specific techniques |
|---|---|---|
| higher energy physics | quantum theory | computer simulation |
| atomic physics | molecular dynamics | spectroscopy |
| nuclear physics | atoms | Tokamak devices |
| molecular physics | electron | thin film |
| condensed matter physics | photons | |
| solid state physics | ions | |
| | electric fields | |
| | computational fluid dynamics | |
| | magnetic fields | |
| | density functional theory | |
| | quantum chemistry | |
| | plasmas | |
| | hydrogen | |
| | thermodynamics | |
| | potential energy | |
| | polarization | |
| | silicon | |
| | temperature | |
| | neutrons | |
| | electron energy level | |
| | ionization | |
| | elementary particle | |
| | ground state | |
| | Reynolds numbers | |
| | photons | |
| | heat transfer | |
| | resonance | |
| | anisotropy | |

The analysis also suggests that the evolution of experimental physics, considering the highest occurrence of keywords in scientific products, is due to three main drivers: research fields, key sub-fields of research and new technologies/specific methods as represented in table 6. However, some domains of research here can be categorized in more than one set and/or can be located at the intersection of two or more sets (Table 6).

In particular, the categorization in table 6 reveals that the rapid development of computer technologies has supported computer simulation, which has many applications in experimental physics driving scholars in new fields of scientific investigations, such as *molecular dynamics* that applies computer simulation methods for studying the physical movements of atoms and molecules, *computational fluid dynamics* that uses numerical analysis and data structures to analyze and solve problems that involve fluid flows, the development in quantum chemistry methods of the *density functional theory* (DFT) based on a computational quantum mechanical modelling used in physics, chemistry and materials science to investigate atoms, molecules, and the condensed phases (DFT methods, *de facto*, has made it possible for quantum chemistry calculations to reach accuracies comparable to those obtained in experiments for molecules of moderate sizes, etc.).

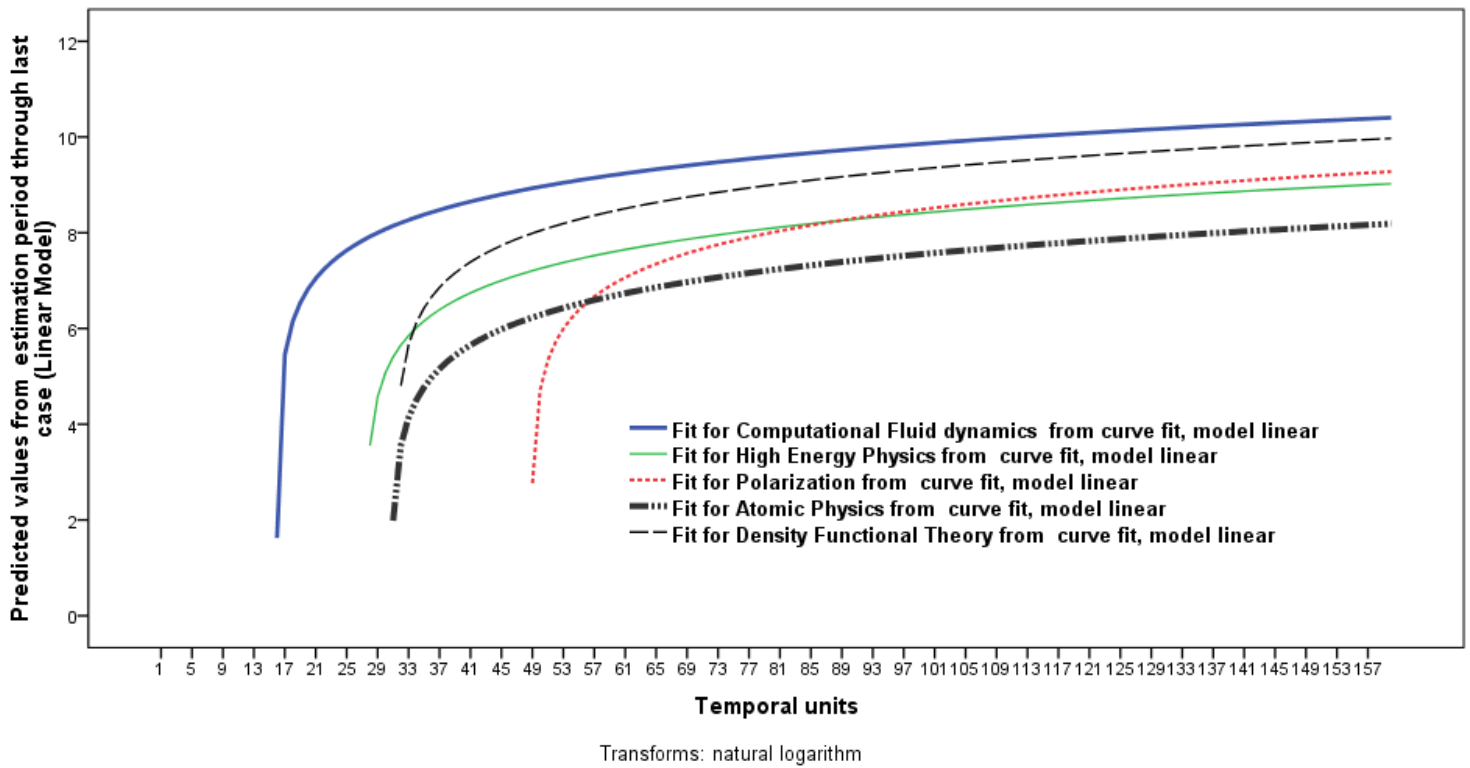

**Figure 5.** Scientific forecasting of driving research fields in experimental physics using sequence plot with predicted values from curve fit of linear model of regression on y-axis (semi log scale)

Finally, this study proposes a scientific forecasting of driving research fields in experimental physics (figure 5). The sequence chart of predicted values of the scientific production of research fields in experimental physics *post*-1900 reveals that studies based on computational fluid dynamics, density functional theory, high energy physics and polarization are the driving research topics in experimental physics rather than atomic physics.

**Discussion, Limitations and Conclusions**

Seidman (1987, p. 131) argues that: "Science is a mode of constructing reality in that like other symbolic constructions of the world (e.g., political ideologies, religion, aesthetic and philosophical theories) it elaborates totalizing symbolic frameworks anchored in broad philosophical theories, moral, and political views about human nature, social order, and historical development. …. Theories, in other words, become part of the cultural symbolism and meanings of a society; they orient and justify action; form elements of our personal and collective identity; and legitimate institutions and public policy. Viewing science in this manner suggests a comparable shift in our understanding of the dynamic of schools". Coccia (2019) claims that science and scientific research are driven by an organized social effort that inevitably reflects the concerns and interests of nations to achieve technical advances and discoveries to take advantage of important opportunities or to cope with environmental threats.

The evolution of science and research fields is due to a cumulative change based on exploration and solution of new and consequential problems in nature and society (cf., Coccia, 2016; 2017a; Scharnhorst et al., 2012; Popper, 1959). Moreover, the dynamics of science tend to follow a process that branches in different disciplines and research fields within and between basic and applied sciences (Mulkay, 1975). In particular, the evolution of scientific fields can be driven by convergence between applied and theoretical sciences (Coccia and Wang, 2016), new scientific paradigms (Kuhn, 1962), new research programmes (Lakatos, 1978), new technologies and breakthrough innovations (Coccia, 2016, 2017, 2017c), fractionalization and specialization of general disciplines, etc. (Coccia, 2018; Crane, 1972; De Solla Price, 1986; Dogan and Pahre, 1990; Mulkay, 1975; van Raan, 2000). Sun et al. (2013, p. 3) show: "the correspondence between the social dynamics of scholar communities and the evolution of scientific disciplines". In general, the evolution of research fields is a natural process of the dynamics of science guided also by curiosity, self-determination and motivation of scholars to explore the unknown in a context of social interactions between scientists, research institutions and countries in an international network of research collaborations (Adams, 2012, 2013; Coccia, 2018, 2019c; Coccia and Bozeman, 2016; Coccia and Wang, 2016; Gibbons et al., 1994; Newman, 2001, 2004; Pan et al., 2012).

The analysis here suggests some empirical results for clarifying the question stated in Introduction, suggesting the characteristics and properties of endogenous processes of the evolution of experimental physics that can be generalized to explain the relationships underlying the evolutionary dynamics of research fields in applied sciences over time and space. In particular, empirical properties of endogenous processes of the evolution of research fields, considering the uniformity and unity found among deeper elements in the system of applied sciences are (cf., Haskins, 1965; Science, 1965; Rousmaniere, 1909; Wassermann, 1989):

[1] *Property of scientific fission*: the division of a research field over time into more research fields that evolve as autonomous entities, generating consequential scientific fissions. The scientific fission of major disciplines is based on processes of specialization, diversification and fractionalization in new research fields. This characteristic also generates the convergence in the long run of research fields into other disciplines, supporting interdisciplinarity and cross-fertilization between applied and theoretical sciences for exploring new directions in science.

*Evidence*. The scientific fission of experimental physics has produced multiple research fields that evolve autonomously in science, generating consequential scientific fissions, such as from astronomy to radio astronomy in 1932, extragalactic astronomy, cosmology, etc.; from radio astronomy to studies of quasars in 1950-1963, pulsars in 1967, etc. (cf., Fig. 6; Mulkay, 1975, p. 518*ff*). Sun et al. (2013) argue that models of science dynamics have attributed the evolution of fields to branching, caused by new discoveries or processes of specialization and fragmentation (cf., Mulkay, 1975; Dogan and Pahre, 1990; Noyons and van Raan, 1998). These models point to the self-organizing development of science exhibiting growth and emergent behavior (cf.,

van Raan, 2000). Other approaches explain scientific progress of new research fields with the synthesis of elements of preexisting disciplines, such as in quantum computing. In this context, Small (1999, p. 812) argues that: "the location of a field can occasionally defy its disciplinary origins". Sun et al. (2013, original emphasis) also claim that: "new scientific fields emerge from *splitting* and merging of … social communities. Splitting can account for branching mechanisms such as specialization and fragmentation, while merging can capture the synthesis of new fields from old ones. The birth and evolution of disciplines is thus guided mainly by the social interactions among scientists".

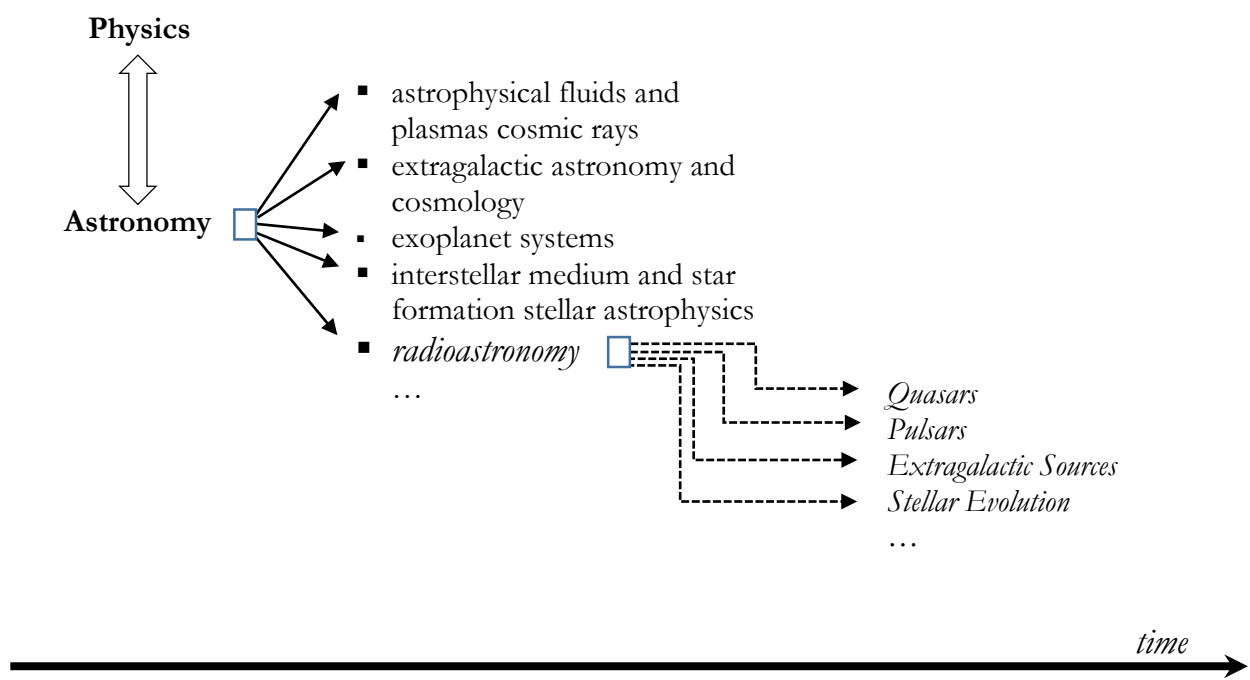

**Figure 6.** Consequential scientific fissions from physics-astronomy, radio astronomy to studies of quasars and other exotic objects in space

In addition, sources of new research fields can be also due the formation of new social groups of scientists migrated from other research fields (Bettencourt et al., 2009; Crane, 1972; Guimera et al., 2005; Wagner, 2008). Hence, evolutionary pathways in science generate new research fields originated from a process of convergence between disciplines, from a specialization within applied or basic sciences or through the combination of multiple disciplines (cf., Coccia and Wang, 2016; Jamali and Nicholas, 2010; Jeffrey, 2003; Riesch, 2014; van Raan, 2000). Sun et al. (2013) state that social interaction among groups of scientists is: "the driving force behind the evolution of disciplines" (cf., Wuchty et al., 2007). In the evolution of scientific fields, Small (1999, p. 812) shows that: "crossover fields are frequently encountered." Hence, interdisciplinarity in science can generate new discoveries and disciplines that support the development of different research fields (cf., Tijssen, 2010).

[2] *Property of ambidextrous drivers of science:* the scientific development is due to ambidextrous driving forces given by scientific discoveries or new technologies. These two drivers have an interaction over the course of time that generates a cross-fertilization supporting sequential scientific and technological change. The overall pattern of scientific development is a complex net of communication of scientific information and technology transfer paths linking together scientific and technological domains. In short, the process of scientific development represents the confluence of new scientific knowledge and technological capabilities within the framework of different scientific fields generating convergence of sciences and scientific fissions inducing new scientific directions.

*Evidence.* For instance, the scientific discovery the human microbiome has created a new research field focused on microbiome research with subsequent evolution of microbial ecology and biomedical sciences (Relman, 2002; Coccia, 2018). Another driver of scientific development is new technologies that destroy existing paradigms creating new conceptual scheme, such as transmission electron microscopy that has supported the discovery of quasicrystals and the evolution of crystallography, metallurgy, materials science, the research field of condensed matter and semiconductor (Shechtman et al. 1984; cf., Coccia, 2016).

[3] *Property of high rates of growth in new driving research fields*: new research fields have higher rates of growth than old ones, providing new directions to the paths of scientific development driven by new scientific schools/communities also fueled by migration of scholars in new fields that have room for new findings, new discoveries, and as a consequence paradigm shifts supporting the general evolution of science.

*Evidence*. The first four on five research fields in experimental physics with the highest scientific production are emerged *post*-1900: computer simulation in 1948, high energy physics in 1929, atomic physics in 1917 and nuclear physics also in 1917. This result suggests that the evolution of experimental physics is driven mainly by new research fields. In particular, 15 research fields supporting experimental physics are emerged after 1900 and have a total number of scientific products equal to 46% on a total of 40 research fields. These new research fields have higher rates of growth than old ones, assuming the characteristic of driving forces of experimental physics. Coccia (2018) also claims that critical research fields can be the driving force of disciplines, providing scientific guideposts that lay out certain definite paths of development. In particular, Coccia (2018) shows that the evolution of a research field is driven by few disciplines that generate more than 80% of documents (*concentration of scientific production in few disciplines*).

[4] *Property of life cycle of research fields*: new research topics emerge in average every 3.9 years with a high creativity rooted-in-war or potential conflicts; moreover, the upwave of scientific cycle, based on scientific production, is about 80 years that is almost the period of one generation of scholars.

*Evidence*. Findings suggest that after 1900 in experimental physics, new research topics emerge in average every 3.9 years, or every decade it emerges in average 2.5 new concepts. New research topics in experimental physics are detected with the first year of the scientific product that includes the keyword/research topic (cf., Fig. 2 and Tab. 3). Moreover, results reveal a cluster of new research topics between WWI and WWII and during 1870s. Warfare is a condition that affects all orders of society life (Coccia, 2015; Stein and Russett, 1980). Although war has many negative effects, it seems to have a crucial permanent connection with the progress of science and technology, driven by new discoveries and technologies originated to solve overriding and relevant problems in the presence of environmental threats (Goldstein, 2003, p. 215; cf., Coccia, 2015; 2018a; 2019a). Stein and Russett (1980) argue that war can propel economic and social change. Wars support high investments in R&D that foster the origin and diffusion of new discoveries, radical and incremental innovations (cf., Clark, 1987; Coccia, 2018a; Constant, 2000). In fact, the innovative and creative spirit is intensified in the presence of effective and/or potential environmental threats associated with wars (Coccia, 2018a, p. 292; Coccia, 2015). In general, war generates demand-side and supply-side effects on socioeconomic systems. The demand-side effects of wars spur a huge demand shock that is due to a massive increase in deficit spending and expansionary policy of nations, supporting investment in Research & Development (R&D) and human resources (cf., Field, 2008). The demand effects, during wars, are coupled to powerful supply-side effects: i.e., learning by doing in military production, spin-off and spillover of scientific breakthroughs from military R&D for solving overriding problems in society (Coccia, 2015; Gemery and Hogendorn, 1993). These joint effects of conflicts can generate a substantial impact on national output, productivity, and as a consequence on scientific, technological and economic growth of nations (cf., Ruttan, 2006; Field, 2008; Coccia, 2019). Results also suggest that average period of the upwave of scientific cycle, based on scientific production, is about 80 years. Moreover, results reveal that emerging research fields in experimental physics, with the highest rate of growth are computational fluid dynamics started in 1966, density functional theory started in 1939, and condensed matter physics started in 1952. These newest research fields in experimental physics, originated during the 1940s, have an average age of 65 years (in 2019$y$) and are driven mainly by a rapid development of computer technologies and computational science.

[5] *Property of science driven by development of general purpose technologies*, the evolution of experimental physics and applied sciences is due to the convergence of experimental and theoretical branches of physics associated with the development of computer, information systems and applied computational science (e.g., computer simulation).

*Evidence*. The rapid development of computer technologies and applied computational science has supported computer simulation, which has wide range of application domains in experimental physics, such as molecular dynamics that applies computer simulation methods for studying the physical movements of atoms and molecules, computational fluid dynamics that uses numerical analysis and data structures to analyze and solve problems that involve fluid flows, the density functional theory based on a computational quantum mechanical modelling used in physics, chemistry and materials science to investigate atoms, molecules, and the condensed phases, etc.

In addition, the evolution of experimental physics seems to be due to three forces given by: *vital research fields, critical sub-domains of research and new technologies/specific techniques. The first* driving force is formed by higher energy physics, atomic physics, nuclear physics, molecular physics, condensed matter physics and solid state physics; *the second one* is quantum theory, molecular dynamics, computational fluid dynamics, quantum chemistry, density functional theory, Reynolds numbers, etc.; finally, *the third force* is given by computer simulation, spectroscopy, Tokamak devices, etc. In general, the underlying drivers of experimental physics are due to all experimental and theoretical aspects of branches in physics that use more and more computer simulation methods.

This study focuses on endogenous processes of the evolution of experimental physics in applied sciences but it would be elusive to limit the evolution of scientific fields to these endogenous factors because the dynamics of science is also due to manifold exogenous factors, such as social contexts of nations, economic growth, democratization of nations, military and political tensions between superpowers to prove scientific and technological superiority, new challenges between superpowers for sustaining global leadership and other events to science (cf., Coccia, 2010, 2011, 2015, 2017; 2018b; 2019, 2019b). As a matter of fact, the evolution of scientific fields, like experimental physics and other applied/basic sciences, is due to expanding human life-interests whose increasing realization constitutes progress that characterizes the human nature for millennia (Coccia and Bellitto, 2018).

Overall, then, this study reveals empirical results, based on the evolution of experimental physics, that may explain and generalize, whenever possible some characteristics of the evolution of scientific fields in applied sciences. These findings can also support best practices of research policy for guiding R&D funding towards new fields that are likeliest to evolve rapidly for maximizing progress of science in society, such as computational fluid dynamics, density functional theory, quantum computing, quantum chemistry, condensed matter physics, etc. However, these conclusions are of course tentative because we know that other things are not equal in the dynamics of science over time and space. To conclude, the inductive study here cannot be enough to explain the comprehensive characteristics of the evolution of research fields and of science, because it is focused on a specific discipline in applied sciences, i.e. the experimental physics, and scientific fields change their scientific borders during the evolution of science and technology. Therefore, the identification of general patterns of the evolution of science and scientific fields in basic and applied sciences—at the intersection of economic, social, psychological, anthropological, philosophical, and biological characteristics of human being—is a non-trivial exercise. The future development of this study is to reinforce proposed results with additional empirical research within other domains of science to provide comprehensive properties that can explain and predict the evolution of different research fields in applied/basic sciences that is important, very important for understanding how foster fruitful scientific trajectories for human progress and wellbeing in future society.